\documentclass[runningheads]{llncs}
\usepackage{graphicx}
\usepackage{amsmath}
\usepackage{amssymb}
\usepackage{multirow}
\usepackage{paracol}
\usepackage{algpseudocode}
\usepackage{caption}
\usepackage{enumitem}
\usepackage{booktabs}
\usepackage{marvosym}
\usepackage[detect-all]{siunitx}

\begin{document}
\title{DDMM-Synth: A Denoising Diffusion Model for Cross-modal Medical Image Synthesis with Sparse-view Measurement Embedding}
\titlerunning{DDMM-Synth}
%
\author{Xiaoyue Li\inst{1} \and Kai Shang\inst{1} \and
Gaoang Wang\inst{2}\textsuperscript{\Letter} \and
Mark D.\ Butala\inst{1}\textsuperscript{\Letter}}
\authorrunning{Li. et al.}
%
\institute{Zhejiang University-University of Illinois Urbana-Champaign
Institute, and College of Information Science and Electronics, Zhejiang University, China \email{markbutala@intl.zju.edu.cn} \and Zhejiang University-University of Illinois Urbana-Champaign
Institute, and College of Computer Science and Technology,
Zhejiang University, Zhejiang University, China \email{gaoangwang@intl.zju.edu.cn}}


\maketitle              
\begin{abstract}
Reducing the radiation dose in computed tomography (CT) is important to mitigate radiation-induced risks. One option is to employ a well-trained model to compensate for incomplete information and map sparse-view measurements to the CT reconstruction. However, reconstruction from sparsely sampled measurements is insufficient to uniquely characterize an object in CT, and a learned prior model may be inadequate for unencountered cases. Medical modal translation from magnetic resonance imaging (MRI) to CT is an alternative but may introduce incorrect information into the synthesized CT images in addition to the fact that there exists no explicit transformation describing their relationship. To address these issues, we propose a novel framework called the denoising diffusion model for medical image synthesis (DDMM-Synth) to close the performance gaps described above. This framework combines an MRI-guided diffusion model with a new CT measurement embedding reverse sampling scheme. Specifically, the null-space content of the one-step denoising result is refined by the MRI-guided data distribution prior, and its range-space component derived from an explicit operator matrix and the sparse-view CT measurements is directly integrated into the inference stage. DDMM-Synth can adjust the projection number of CT a posteriori for a particular clinical application and its modified version can even improve the results significantly for noisy cases. Our results show that DDMM-Synth outperforms other state-of-the-art supervised-learning-based baselines under fair experimental conditions. 
\end{abstract}
\section{Introduction}

\begin{figure}[!t]
\centering
\includegraphics[width=\textwidth]{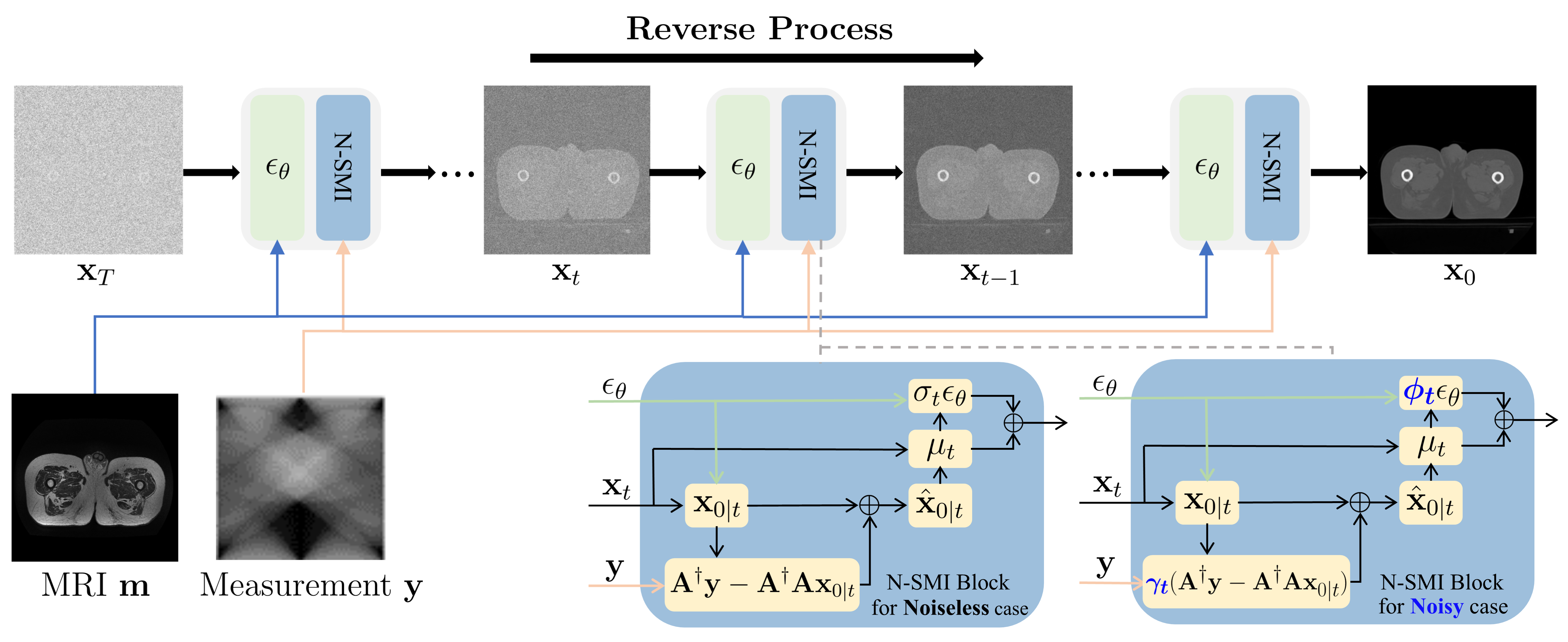}
\caption{High-level illustration of the reverse sampling process for DDMM-Synth. The green block is the network $\epsilon_\theta$, which is conditioned by MRI $\mathbf{m}$. The blue block is the null-space measurement inference (N-SMI) block, where the measurement matrix $\mathbf{A}$ and measurement $\mathbf{y}$ are used to guide the inference process. Here we show two N-SMI blocks: one for the case with noise and the other without.} \label{fig_process_flow}
\end{figure}

Computed tomography (CT) plays an essential role in medical diagnosis. However, obtaining CT measurements with complete projections, i.e., the sinogram, can potentially lead to dangerous radiation exposure, threatening the health of patients, especially for individuals with chronic conditions or undergoing cancer treatment. Reducing the radiation dose is vital but unavoidably increases artifacts in CT images, which could compromise a clinical diagnosis.\\
\indent The advent of data-driven methods offers fresh means to mitigate these aforementioned issues. Some prior studies have investigated the reconstruction of CT images from sparse-view measurements by learning to directly obtain CT reconstructions from partial measurements \cite{zhu2018image,shen2019patient,wurfl2018deep,wei20202}. Nevertheless, most of these are supervised learning techniques and need re-train a dedicated model when the measurement process changes, which lacks generalizability and flexibility. Moreover, for ill-posed problems like CT, sparsely sampled measurements are insufficient to uniquely specify an object. Medical image translation is another approach to solve the above problem \cite{han2017mr,chartsias2017adversarial,armanious2020medgan,bahrami2020new} which generally operates by determining a mapping from the source modality to the target modality, usually with specific focus on the translation from Magnetic Resonance Imaging (MRI) to CT. Since there exists no explicit and reliable relationship between them and sometimes MRI exhibit occluded sections  (see, e.g, the left-most image in the second row of Fig.~\ref{fig2}), image translation with only MRI images opens the possibility to introduce unreliable information into the synthesized CT images, which affects the diagnostic quality. Recently, the diffusion model \cite{song2019generative,ho2020denoising,song2020score,song2021scorebased} has emerged as a competitive candidate to realize high-fidelity synthetic results and has attracted the attention of medical imaging researchers \cite{jalal2021robust,peng2022towards,chung2022score,song2022solving}. Diffusion model consists of a forward stage that gradually injects Gaussian noise into a sample over sufficient steps and a reverse stage to iteratively denoise and finally recover the noiseless image.
Diffusion models have the advantage of analytical explainability, utilizing an optimal transport process that results in high-appearance similarity \cite{khrulkov2022understanding}, consistently achieving state-of-the-art performance in numerous image synthesis tasks \cite{croitoru2022diffusion}. \\
\indent In this paper, we introduce a new methodology using a diffusion model guided by both MRI and sampled prior CT information embedding for medical image synthesis, termed DDMM-Synth, which is designed as an attempt to deal with some clinical diagnoses, e.g., in evaluating traumatic pelvic injury \cite{smith2019advanced,nuchtern2015significance}, which need depend on multiple imaging modalities, aiming to obtain more realistic and reliable medical image synthesis while reducing the radiation risk to patients. In the DDMM-Synth framework, a pretrained diffusion model conditioned on MRI was utilized to provide valuable prior knowledge for CT reconstruction. Inspired by the great potential of reverse sampling via a range-null space decomposition \cite{schwab2019deep,wang2022gan,wang2022zero}, we then introduce the null-space measurement inference (N-SMI) block used to insert the refined null-space contents and necessary scanning information from sparse-view CT measurement into the reverse inference process (see Fig.~\ref{fig_process_flow}) without additional training. Intuitively, DDMM-Synth introduces guidance in the denoising process by manipulating the denoising object at each step with the known measurements, transforming both by the pseudo-inverse of the measurement matrix. \\
\indent Our main contributions are as follows: (1) We propose a novel diffusion model that integrates both an implicit data distribution prior mapping from MRI to CT images and effective information derived from sparse sampled CT measurements. This integrated approach enables higher-fidelity synthesis of CT images, which better reflects the structural and anatomical details of low-dose generated CT images. (2) Our method does not require a fixed measurement and can be adapted to any particular measurement scenario as long as the mapping from images to measurements is linear, which is generally the case in clinical applications. (3) Different from existing measurement-conditioned diffusion models which only work on noiseless measurements, our modified method DDMM-Synth-noise improves the synthesized results significantly for noisy measurements, enabling the use of our model even in challenging conditions.

\section{DDMM-Synth}

DDMM-Synth aims to synergistically combine MRI tissue information and a sparse sampled CT reconstruction to yield synthetic CT images that preserve the detailed tissue features encoded in the MRI while maintaining data consistency with CT images. Our model employs a conditional denoising diffusion model and a null-space measurement decomposition-based reverse sampling process. The core principles of DDMM-Synth are detailed below.

\subsubsection{Conditional Denoising Process}
Technically, the reverse denoising process can be defined as the posterior distribution $p\left(\mathbf{x}_{t-1} \mid \mathbf{x}_t, \mathbf{x}_0\right)$ with mean ${\mu}_t\left(\mathbf{x}_t,\mathbf{x}_0\right)=\dfrac{1}{\sqrt{\alpha_t}}\left(\mathbf{x}_t-\epsilon\dfrac{1-\alpha_t}{\sqrt{\bar{\alpha}_t}}\right)$ and variance $\sigma^2_t=\dfrac{1-\bar{\alpha}_{t-1}}{1-\bar{\alpha}_t}\beta_t$ to predict the noiseless objects $\mathbf{x}_0 \sim q\left(\mathbf{x}\right)$ from random Gaussian noise $\mathbf{x}_T$ from a $T$-step forward process, where $\alpha_t = 1 - \beta_t$ and $\bar{\alpha}_t=\prod\limits_{i=1}^T\alpha_i$, $\beta_t \in (0, 1)$ is a predefined value. $q\left(\mathbf{x}\right)$ means the data distribution for the training dataset. The network $\epsilon_\theta\left(\mathbf{x}_{t-1},\mathbf{x}_t\right)$ is trained aiming to predict the only uncertain variable $\epsilon$ for each reverse step $t$ and then provide the estimated mean $\mu_\theta(\mathbf{x}_t, t)$ to obtain $\mathbf{x}_{t-1}$. 


Following the conditional model utilized in \cite{saharia2022image}, the first step of DDMM-Synth is to adapt the diffusion model to be  conditional by concatenating each MRI image with a corresponding input noisy CT image when training the network $\epsilon_\theta$. Guided by the MRI images $\mathbf{m}$, the objective can be modified as follows:
\begin{equation}
    \mathbf{L}_{t-1}^{\text{conditional}}=\mathbb{E}_{\mathbf{x}_0, \epsilon \sim \mathcal{N}(\mathbf{0}, \mathbf{I})}\left[\left \| \epsilon-\epsilon_\theta \left(\sqrt{\bar{\alpha}_t}\mathbf{x}_0+\sqrt{1-\bar{\alpha}_t} \epsilon, \mathbf{m}, t\right)\right \|^2\right].
\end{equation}

\subsubsection{Inference via Range-Null Space Decomposed Medical Measurement}


The standard forward model in CT is given by
\begin{equation}
    \mathbf{y}=\mathbf{A}\mathbf{x}+\mathbf{n},
\end{equation}
where $\mathbf{y}$ denotes the measurements, $\mathbf{A}$ is the linear measurement operator, $\mathbf{x}$ are the discretized, unknown densities to be reconstructed, and $\mathbf{n}$ represents acquisition noise. \\
\indent Traditional CT can generally be summarized as a trade-off between a data-fidelity term and an image-prior term. The former models the physical process to ensure data consistency while the latter is used to regularize the reconstructed images via hand-crafted priors. However, traditional techniques cannot faithfully reconstruct realistic details and compromise by smoothing the results. Moreover, it is challenging to adjust the regularization to balance the data consistency and the priors.\\
\indent From the perspective of the range-null space decomposition \cite{schwab2019deep,wang2022gan,wang2022zero}, data consistency is exclusively linked to range-space content and can be accurately computed. Thus, the fundamental challenge lies in identifying the appropriate null-space content that should be consistent with prior information to yield high-fidelity results. Neglecting acquisition noise, the measurement process can be decomposed as
\begin{equation}
    \mathbf{A}\mathbf{x} \equiv \mathbf{A}\mathbf{A}^\dagger \mathbf{A} \mathbf{x} + \mathbf{A}(\mathbf{I} - \mathbf{A}^\dagger \mathbf{A}) \mathbf{x} \equiv \mathbf{A} \mathbf{x} + \mathbf{0} \equiv \mathbf{y},
\end{equation}
where $\mathbf{A}^\dagger$ is the pseudo-inverse matrix that satisfies $\mathbf{A}\mathbf{A}^\dagger\mathbf{A} \equiv \mathbf{A}$, $\mathbf{A}^\dagger \mathbf{A} \mathbf{x}$ is the portion in the range-space of $\mathbf{A}$, $(\mathbf{I} - \mathbf{A}^\dagger \mathbf{A})\mathbf{x}$ is the portion in the null-space.

For ill-posed problems like CT, there exist infinite $\hat{\mathbf{x}}$ that satisfy the constraint $\mathbf{A}\hat{\mathbf{x}}=\mathbf{y}$. The solution $\hat{\mathbf{x}}$ can be derived as $\hat{\mathbf{x}}=\mathbf{A}^\dagger\mathbf{y}+(\mathbf{I}-\mathbf{A}^\dagger\mathbf{A})\bar{\mathbf{x}}$, where $\mathbf{A}^\dagger\mathbf{y}$ aims to satisfy the data consistency and $\bar{\mathbf{x}}$ determines the data distribution priors. On account of the outstanding abilities to fit data distributions with diffusion models, here we replace the null-space portion of one-step denoising result during the inference process, aiming for a higher-fidelity synthetic result that also satisfies the original data distribution prior mapping from the corresponding MRI guidance. The reverse inference process of the diffusion model can be modified as follows:
\begin{gather}
    p\left(\mathbf{x}_{t-1} \mid \mathbf{x}_t,\mathbf{x}_0\right):=\mathcal{N}\left(\mathbf{x}_{t-1};\mu_t(\mathbf{x}_t,\hat{\mathbf{x}}_{0 \mid t}),\sigma^2_t\mathbf{I}\right),\\
    \hat{\mathbf{x}}_{0 \mid t}=\mathbf{A}^\dagger\mathbf{y}+(\mathbf{I}-\mathbf{A}^\dagger\mathbf{A})\mathbf{x}_{0 \mid t}=\mathbf{x}_{0 \mid t}-\mathbf{A}^\dagger(\mathbf{A}\mathbf{x}_{0 \mid t}-\mathbf{y}),\label{eq.x0t}
\end{gather}
where $\mathbf{x}_{0 \mid t}$ can be derived from $\mathbf{x}_t$ and the noise model $\epsilon_\theta$ as follows:
\begin{equation}
    \mathbf{x}_{0 \mid t}=\dfrac{1}{\sqrt{\bar{\alpha_t}}}\left(\mathbf{x}_t-\epsilon_\theta(\mathbf{x}_t,\mathbf{m}, t)\sqrt{1-\bar{\alpha}_t}\right).
\end{equation}
Then, we employ a step-by-step reverse sampling with $p(\mathbf{x}_{t-1} \mid \mathbf{x}_t,\hat{\mathbf{x}}_{0 \mid t})$ and finally obtain the denoised CT image.\\
\indent Importantly, the range-null space decomposed measurement can be further extended to the noisy case by modifying Eq.~\ref{eq.x0t} as:
\begin{equation}
   \hat{\mathbf{x}}_{0 \mid t} = \mathbf{A}^\dagger\mathbf{y}+(\mathbf{I}-\mathbf{A}^\dagger\mathbf{A})\mathbf{x}_{0 \mid t}\notag \\ 
   = \mathbf{x}_{0 \mid t} - \mathbf{A}^\dagger(\mathbf{A}\mathbf{x}_{0 \mid t} - \mathbf{A}\mathbf{x})+\mathbf{A}^\dagger\mathbf{n}, \quad \mathbf{A}^\dagger\mathbf{n} \sim \mathcal{N}(\mathbf{0}, \sigma^2_{\mathbf{n}}\mathbf{I}).
\end{equation}

To account for the noise added into $\hat{\mathbf{x}}_{0 \mid t}$, we define a scaling parameter $\gamma_t$ to adjust the range-space correction. Then, the reverse process can be rewritten as:
\begin{align}
    \hat{\mathbf{x}}_{0 \mid t} &= \mathbf{x}_{0 \mid t} - \gamma_t\mathbf{A}^\dagger(\mathbf{A}\mathbf{x}_{0 \mid t} - \mathbf{A}\mathbf{x} - \mathbf{n}) \\
    &= \mathbf{x}_{0 \mid t} - \gamma_t\mathbf{A}^\dagger(\mathbf{A}\mathbf{x}_{0 \mid t} - \mathbf{A}\mathbf{x}) + \sigma_{\mathbf{n}}\gamma_t\epsilon_{\mathbf{n}}, \quad \epsilon_{\mathbf{n}} \sim \mathcal{N}(\mathbf{0},\mathbf{I}).
\end{align}
Next, the denoising process in the DDPM sampling strategy can be written as:
\begin{equation}
    \mathbf{x}_{t-1}=\frac{\sqrt{\bar{\alpha}_{t-1}}\beta_t}{1-\bar{\alpha}_t}\hat{\mathbf{x}}_{0 \mid t} + \frac{\sqrt{\alpha_t}(1-\bar{\alpha}_{t-1})}{1-\bar{\alpha}_t}\mathbf{x}_t+\sigma_t\epsilon, \quad \epsilon \sim \mathcal{N}(\mathbf{0},\mathbf{I}). \label{denoise_process}
\end{equation}
Given that the total noise level cannot exceed the noise distribution of Eq.~\ref{eq.x0t} in Section~2, i.e., $\mathcal{N}(\mathbf{0},\sigma^2_t\mathbf{I})$, Eq.~\ref{denoise_process} can be expressed as:
\begin{gather}
     \mathbf{x}_{t-1}=\frac{\sqrt{\bar{\alpha}_{t-1}}\beta_t}{1-\bar{\alpha}_t}\bigl[\mathbf{x}_{0 \mid t} - \gamma_t\mathbf{A}^\dagger(\mathbf{A}\mathbf{x}_{0 \mid t} - \mathbf{A}\mathbf{x})\bigr] + \frac{\sqrt{\alpha_t}(1-\bar{\alpha}_{t-1})}{1-\bar{\alpha}_t}\mathbf{x}_t+\epsilon_{measure}+\epsilon_{extra},\\
     \epsilon_{measure} = \frac{\sqrt{\bar{\alpha}_{t-1}}\beta_t}{1-\bar{\alpha}_t}\sigma_{\mathbf{n}}\gamma_t\epsilon_{\mathbf{n}} \sim \mathcal{N}\biggl(\mathbf{0}, \Bigl[\frac{\gamma_t\sqrt{\bar{\alpha}_{t-1}}\beta_t\sigma_{\mathbf{n}}}{1-\bar{\alpha}_t}\Bigr]^2 \mathbf{I}\biggr),\\
     \epsilon_{extra} \sim \mathcal{N}(\mathbf{0},\phi_t\mathbf{I}),\\
     \epsilon = \epsilon_{measure} + \epsilon_{extra} \sim \mathcal{N}\biggl(\mathbf{0},\Bigl[\frac{\gamma_t\sqrt{\bar{\alpha}_{t-1}}\beta_t\sigma_{\mathbf{n}}}{1-\bar{\alpha}_t}\Bigr]^2 + \phi_t\biggr)=\mathcal{N}(\mathbf{0},\sigma^2_t\mathbf{I}).
\end{gather}
Here the total noise in the one-step denoising process can be divided into measurement noise $\epsilon_{measure}$ and extra noise $\epsilon_{extra}$ components, where $\phi_t$ scales the extra noise component in order to agree with the general noise variance. 
To assure the general noise follows the Gaussian distribution $\mathcal{N}(\mathbf{0},\sigma^2_t\mathbf{I})$. Following the two principles proposed by \cite{wang2022zero}, for simplicity, $\phi_t$ and $\gamma_t$ were set as:
\begin{gather}
    \phi_t = \sigma^2_t - \big(\frac{\gamma_t\sqrt{\bar{\alpha}_{t-1}}\beta_t\sigma_{\mathbf{n}}}{1-\bar{\alpha}_t}\bigr)^2,\notag
    \gamma_t= \begin{cases}1, & \sigma_t \geq \frac{\sqrt{\bar{\alpha}_{t-1}}\beta_t\sigma_{\mathbf{n}}}{1-\bar{\alpha}_t} \\  \frac{\sigma_t(1-\bar{\alpha}_t)}{\sqrt{\bar{\alpha}_{t-1}}\beta_t\sigma_{\mathbf{n}}}, & \sigma_t<\frac{\sqrt{\bar{\alpha}_{t-1}}\beta_t\sigma_{\mathbf{n}}}{1-\bar{\alpha}_t}\end{cases}.\notag
\end{gather}


An overview of the reverse sampling process is shown in Fig.~\ref{fig_process_flow}. The N-SMI block employed, either noiseless or noisy, can be selected depending on the severity of the acquisition noise. 

\section{Experiments and Results}
\subsubsection{Experimental Setup}
We evaluate the performance of DDMM-Synth on two public datasets. The first is the Gold Atlas project dataset \cite{nyholm2018mr}, which is a multi-modal pelvic MRI-CT dataset where the CT images have been co-registered to fit the anatomy of paired T2 MRI. Following \cite{boni2020mr}, images from site 2 and site 3 were selected as the training dataset and data from site 1 was used as the test dataset. The first and last three axial slices were removed from each volume to avoid aliasing effects. This scheme can also test the robustness of DDMM-Synth as the data in the test dataset were acquired from a different scanner than the training dataset. We have also tested on the BRATS2018 dataset \cite{menze2014multimodal}. Images from 50 low grade glioma (LGG)
were selected as the training set while another 15 patients are selected as the test dataset. Note that we conducted the task under the assumption of using the T2 MRI as the reference (as the dataset does not include CT) and FLAIR as the target to further confirm the effectiveness of our method. Two metrics were used to evaluate the performance: peak signal-to-noise ratio (PSNR) and structural similarity (SSIM).\\ 
\indent We base the implementation on the SR3 model \cite{saharia2022image}. The total number of epochs was set to 100, the U-Net was trained by the Adam optimizer with a learning rate of $1\times{10}^{-4}$. During the training phase, we set $T = 2000$. Using the DDIM sampler \cite{song2020denoising} to accelerate the reverse process, we reduced the number of time steps from 2000 to 100 steps. We simulate sinograms with a parallel-beam geometry using projection angles uniformly distributed over 180$^\circ$. 

\subsubsection{Baseline Methods}
We compare DDMM-Synth with several state-of-the-art generative models: pix2pix \cite{isola2017image}, pGAN \cite{dar2019image}, medSynth \cite{nie2018medical}, and Self-attention GAN (SAGAN) \cite{zhang2019self}. For fair comparisons among competing methods, the same modified 70$\times$70 patchGAN discriminator is used for all methods. We also modified SAGAN which was originally for unconditional cases by incorporating self-attention modules into the pGAN. In addition, all the competing models are conditioned by concatenating MRI images with corresponding sparse-view CT images with 23 projection angles along the channel dimension.

\begin{table}
\caption{Performance evaluation of the CT synthesis task. Boldface indicates the best results and $N_p$ is the number of projections over 180$^\circ$.}\label{tab1}

\setlength{\tabcolsep}{2.78mm}
\centering
\begin{tabular}{@{}lccccc@{}} 
\toprule
                        &  & \multicolumn{2}{c}{Gold Atlas}                        & \multicolumn{2}{c}{BRATS2018} \\ \cmidrule{3-6} 
  Method                                              & $N_p$                             &  PSNR                      & SSIM                      & PSNR          & SSIM          \\ \midrule
pix2pix \cite{isola2017image}                                      & 23                                               & 28.98$\pm1.87$               & 0.887$\pm0.019$                     & 24.01$\pm1.91$   & 0.797$\pm0.028$         \\ 
pGAN \cite{dar2019image}                                           & 23                                               & 29.19$\pm1.64$                     & 0.893$\pm0.012$                     & 24.16$\pm1.44$ & 0.810$\pm0.027$         \\ 
medSynth \cite{nie2018medical}                                     & 23                                               & 30.03$\pm1.81$                     & 0.891$\pm0.023$                   & 24.32$\pm2.06$ & 0.813$\pm0.031$         \\ 
SAGAN \cite{zhang2019self}                                         & 23                                               & 31.01$\pm2.01$                     & 0.912$\pm0.021$                    & 26.62$\pm1.97$ & 0.826$\pm0.027$         \\ \midrule
DDMM-Synth & 10            & 29.82$\pm2.10$                     & 0.861$\pm0.020$                     & 26.47$\pm1.85$         & 0.796$\pm0.025$         \\ 
           & 20            & 33.12$\pm1.99$                     & 0.936$\pm0.019$                     & 27.03$\pm1.27$         & 0.863$\pm0.021$        \\ 
           & 23            & \pmb{33.79}$\pm1.98$               & \pmb{0.941}$\pm0.019$               & \pmb{27.54}$\pm1.25$   & \pmb{0.872}$\pm0.022$  \\ \bottomrule
\end{tabular}
\end{table}

\subsubsection{Results and Discussion}
\begin{figure}[!t]
\centering
\includegraphics[width=1\textwidth]{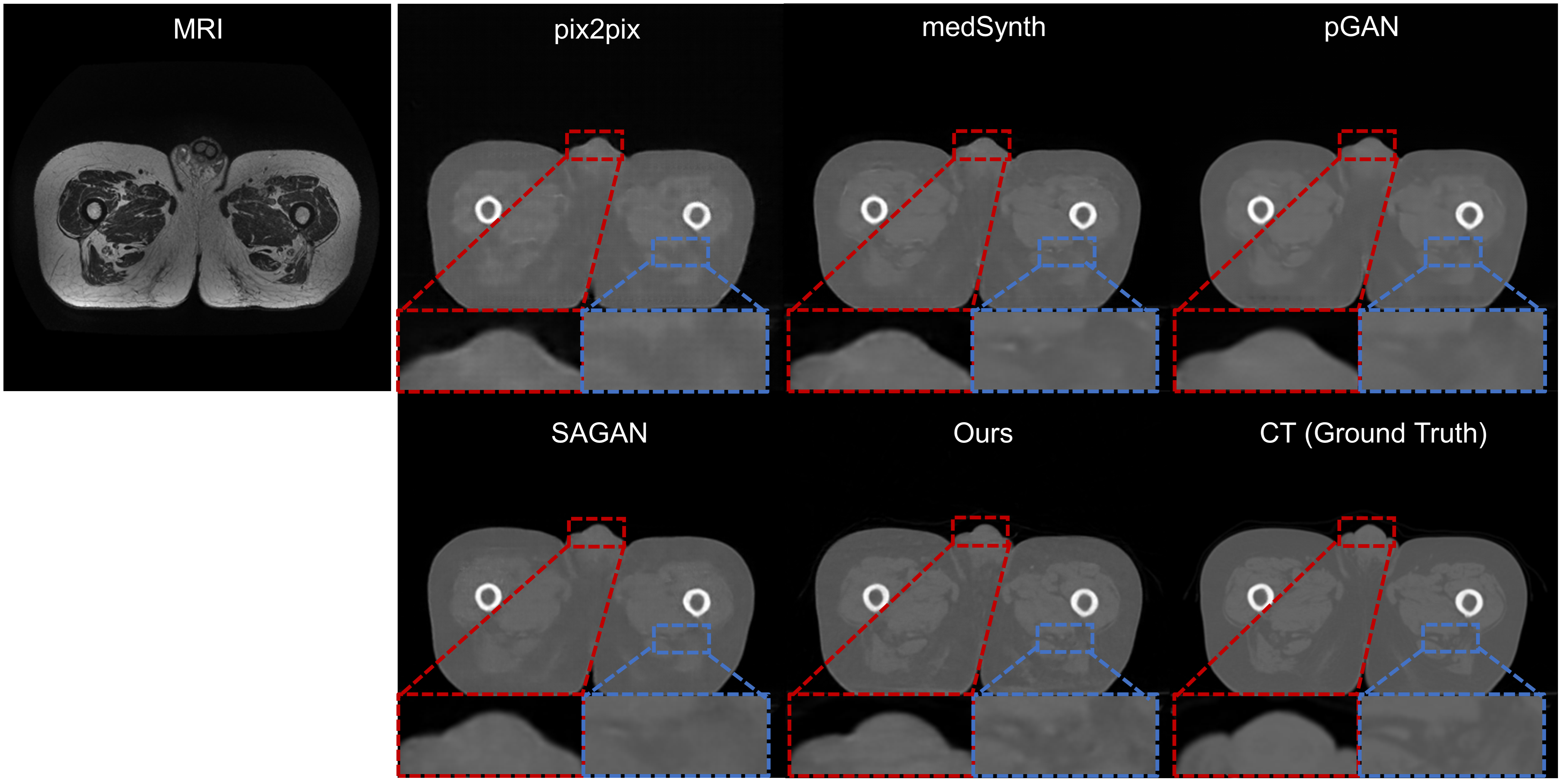}
\caption{DDMM-Synth was compared to other baseline generative methods using the Gold Atlas project dataset.} \label{fig1}
\end{figure}

\begin{figure}[!t]
\centering
\includegraphics[width=1\textwidth]{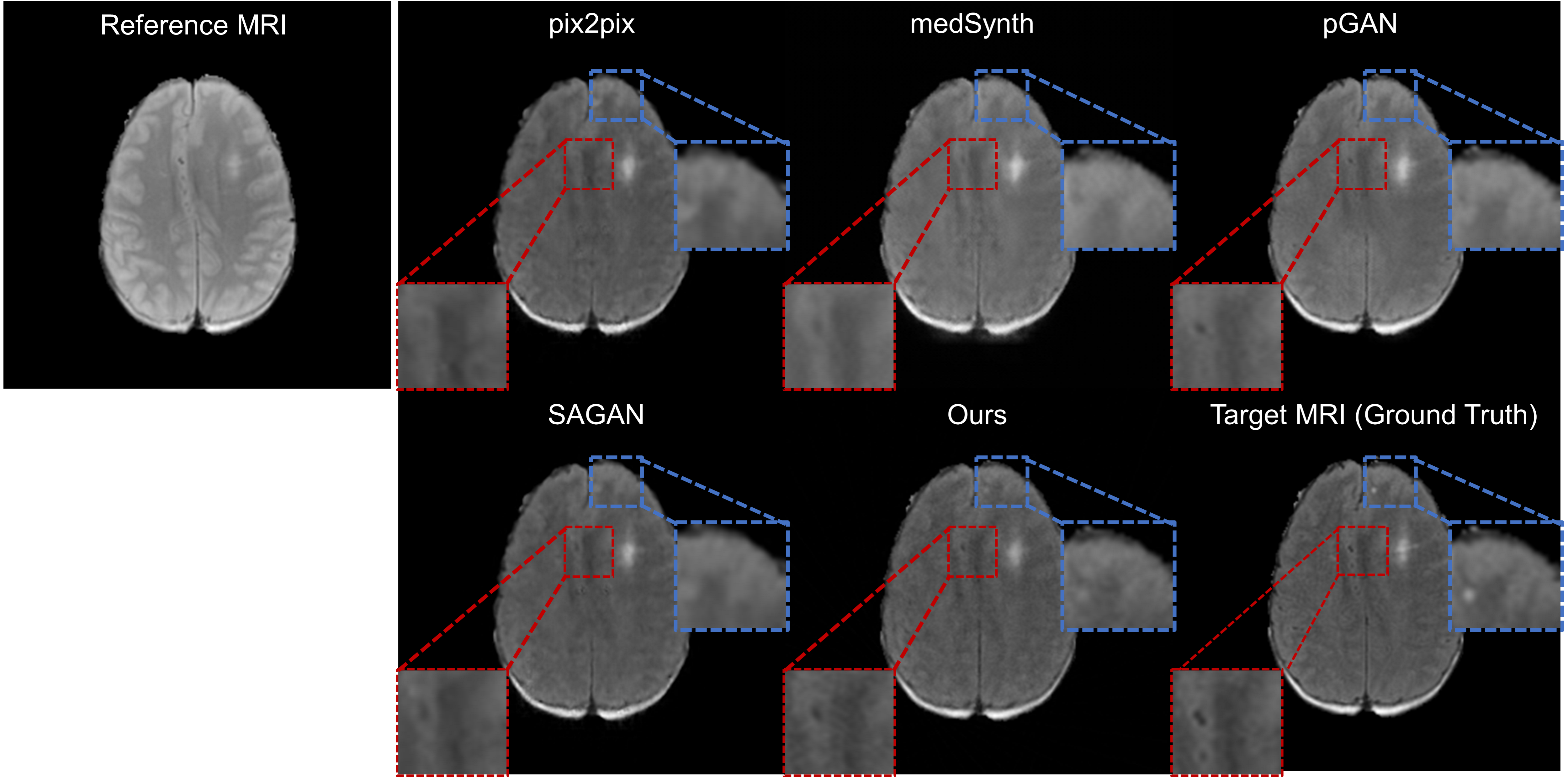}
\caption{DDMM-Synth was compared to other baseline generative methods using the BRATS2018 dataset.} \label{fig2}
\end{figure}

\begin{figure}
\centering
\includegraphics[width=1\textwidth,height=0.220\textwidth]{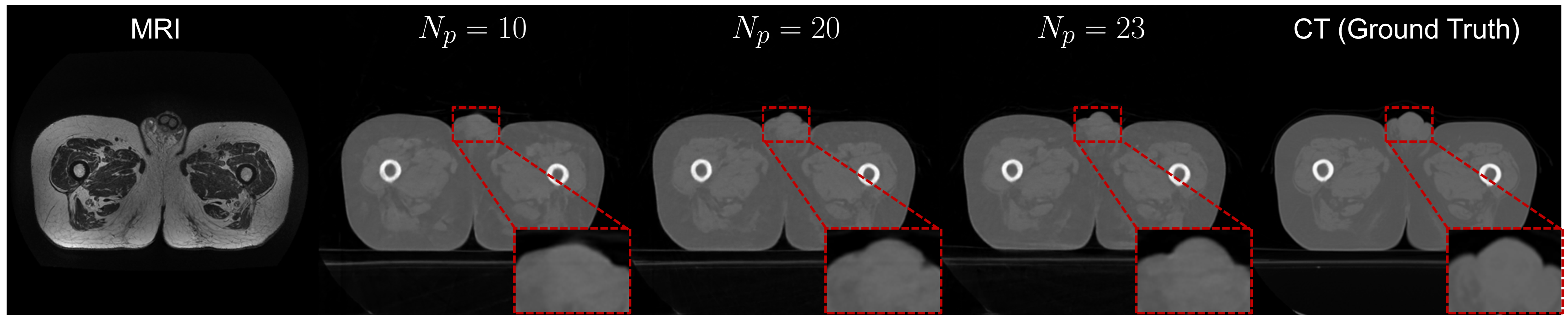}
\caption{Examples of synthetic CT results with different numbers of projections on the Gold Atlas project dataset. Here we show the results with 10, 20 and 23 projections, respectively, over 180$^\circ$.} \label{fig3}
\end{figure}

\begin{figure}[!t]
\centering
\includegraphics[width=1\textwidth,height=0.260\textwidth]{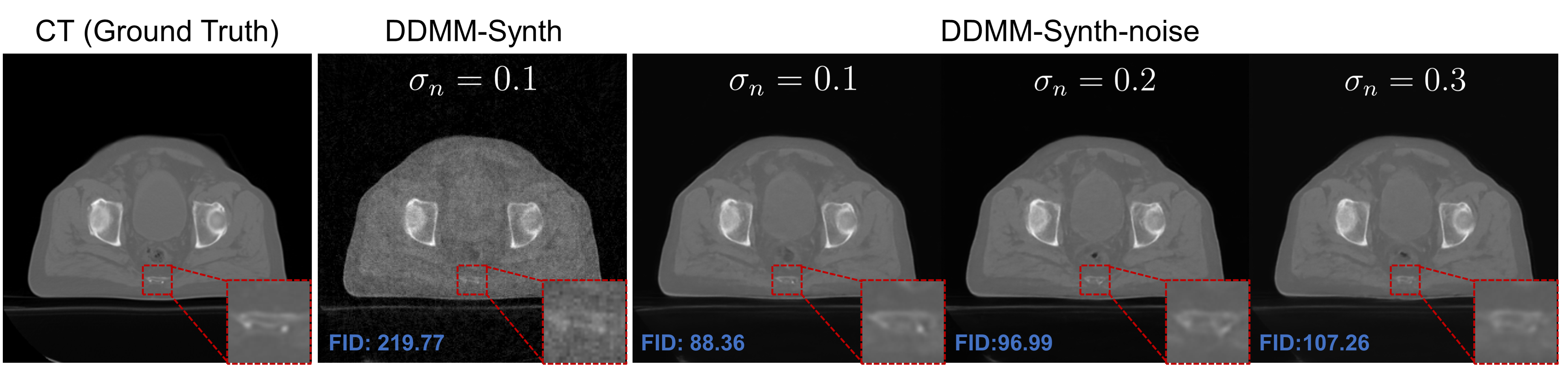}
\caption{DDMM-Synth-noise was able to solve noisy cases with $\mathbf{\sigma}_n$ equal to 0.1, 0.2, and 0.3 respectively. FID scores are shown at the bottom to evaluate each case.} \label{noise_case}
\end{figure}

Table~\ref{tab1} lists the evaluation results. In most cases, DDMM-Synth has the best quantitative results compared to the comparison methods. We consider the case of changing the projection number at the reverse sampling stage, the synthetic CT results with different numbers of projections on the Gold Atlas project dataset can be found in Fig.~\ref{fig3}. We found our results with 20 projections are even better than the competing methods using 23 measurements. Representative target images are displayed in Fig.~\ref{fig1} and Fig.~\ref{fig2}. As can be seen, non-attentional GANs suppress noise, but sacrifice certain details and shift local structures, especially as depicted in the magnified region of interest (ROI). Attention-augmented models offer more noticeable performance benefits than non-attentional GANs, but clear artifacts remain. Compared to the baseline methods, DDMM-Synth synthesizes CT images with fewer artifacts and with more reliable structural feature recovery and higher anatomical fidelity, particularly in the ROI. The results emphasize that diffusion-based models currently have the unparalleled ability to fit data distributions and generate high-quality images. We also explore the performance of our model in noisy cases. Since Frechet Inception Distance (FID) \cite{heusel2017gans} performs better in evaluating visible noises, here we replace the metric with FID scores. Fig.~\ref{noise_case} shows the significant denoising improvement of DDMM-Synth-noise in comparison to DDMM-Synth.

\subsubsection{Ablation Study}
\begin{figure}[!t]
\centering
\includegraphics[width=1\textwidth,height=0.35\textwidth]{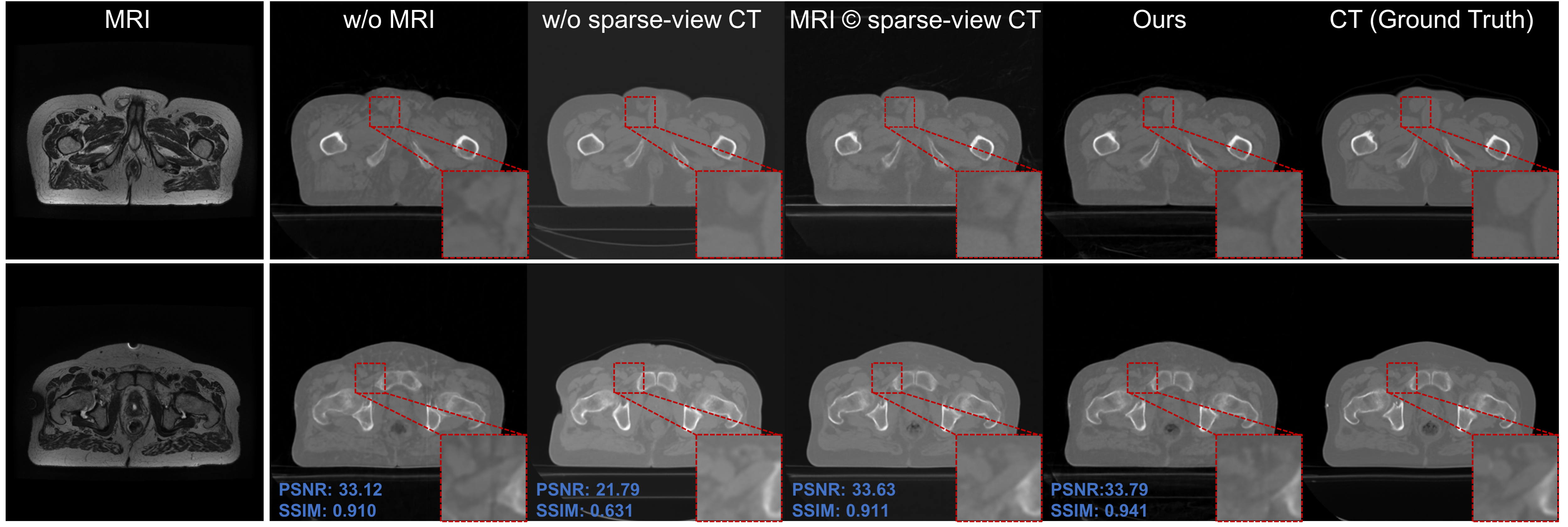}
\caption{Ablation study: DDMM-Synth was compared against three variant models. Here we show the results of two samples from the test dataset.} \label{fig4}
\end{figure}

We conducted a set of ablation studies to quantitatively assess the performance of two conditional elements and the way of inserting conditions into DDMM-Synth.
First, we investigate the benefit of adding sparse-view CT information into DDMM-Synth. The modified CT null-space content was ablated from the reverse sampling stage and the variant method was adjusted to be a pure image translation model with paired MRI images as input. Then, to examine the contribution of MRI images, we removed the conditional guidance (MRI images) during DDMM-Synth training. Third, we investigated how to utilize CT information in DDMM-Synth. As opposed to guiding from the reverse inference stage using a null-space CT measurement, we resorted to directly adding sparse-view CT information into the training phase, which means concatenating sparse-view CT images with paired MRI as input of the noise model $\epsilon_\theta$. Fig.~\ref{fig4} displays the results of the ablation studies on the MRI-CT dataset. The first two columns show the results without MRI and sparse-view CT information embedding respectively, while the third column presents the results generated by the scheme of adding both MRI and CT during training. The evaluation metrics are shown as well. From Fig.~\ref{fig4}, it can be seen that DDMM-Synth images more closely resemble real CT images and exhibit greater spatial acuity when compared to the other three variants. Note that even though the evaluation results of the variant without MRI-guidance perform well, strip artifacts introduced by sparse sampled measurement are apparent. To summarize, the results indicate that MRI can offer more texture detail when mapped to CT images while the embedding of sparse CT prior information ensures greater data consistency. In addition, the guidance mode of DDMM-Synth outperforms the conditional guidance scheme of directly concatenating two conditions.

\section{Conclusion}
In this study, we introduced a novel medical image synthesis approach tailored to MRI to CT translation based on a conditional diffusion model and range-null space decomposition. DDMM-Synth is able to improve the contextual mapping from MRI to CT and while preserving data consistency of CT sparse-view scanning, even for noisy cases. Our model achieves superior synthesis quality compared to recent image translation methods. DDMM-Synth holds great promise for medical image synthesis tasks.


\subsubsection{Limitations and Future Study}

Though our model shows superiority in CT synthesis, we should note the limitations of this method. Since MRI and CT are scanned by different devices, the image registration between them must occur first to ensure the model can accurately map tissue features from MRI to CT images. Inspired by \cite{kong2021breaking}, the addition of an image registration block into the model could provide a mitigation strategy and will be considered in future work. It is also worth noting that the inference of diffusion model is relatively slow. Our future efforts will also focus on improving the inference speed of DDMM-Synth.

\end{document}